\documentclass{kluwer}    

\usepackage[dvips]{graphicx}

\newdisplay{guess}{Conjecture}

\def\lesssim{\,\hbox{\lower0.6ex\hbox{$\sim$}\llap{\raise0.6ex\hbox{$<$}}}\,}
\def\gtsim{\,\hbox{\lower0.6ex\hbox{$\sim$}\llap{\raise0.6ex\hbox{$>$}}}\,}
\def\la{\,\hbox{\lower0.6ex\hbox{$\sim$}\llap{\raise0.6ex\hbox{$<$}}}\,}
\def\ctss{Counts s$^{-1}$}

\begin{document}
\begin{article}
\begin{opening}
\title{Future X-ray timing missions}
\author{Didier \surname{Barret}}
\institute{Centre d'Etude Spatiale des Rayonnements, France}
\author{Michiel \surname{van der Klis}}
\institute{University of Amsterdam, Netherlands}
\author{Gerry K. \surname{Skinner}}
\institute{University of Birmingham, England}
\author{Rudiger \surname{Staubert}}
\institute{University of T\"ubingen, Germany}
\author{Luigi \surname{Stella}}
\institute{Astronomical Observatory of Roma, Italy}
\runningauthor{Didier Barret et al.}
\runningtitle{Future X-ray timing missions}
\date{November 2nd, 2000}
\begin{abstract}
Thanks to the Rossi X-ray Timing Explorer (RXTE), it is now widely
recognized that fast X-ray timing can be used to probe strong gravity
fields around collapsed objects and constrain the equation of state of
dense matter in neutron stars.  We first discuss some of the
outstanding issues which could be solved with an X-ray timing mission
building on the great successes of RXTE and providing an order of
magnitude better sensitivity.  Then we briefly describe the
``Experiment for X-ray timing and Relativistic Astrophysics'' (EXTRA)
recently proposed to the European Space Agency as a follow-up to RXTE
and the related US mission ``Relativistic Astrophysics Explorer''
(RAE).
\end{abstract}
\end{opening}
\section{Introduction}
In the last few years the predicted sub-millisecond plasma orbital
periods and millisecond spin periods of accreting low-magnetic-field
neutron stars (NS) have finally been detected by RXTE. The same
satellite also found 0.1~kHz Quasi-Periodic Oscillations (QPOs) in
black hole candidates (BH) and, the first (and so far only) accreting
millisecond pulsar (see van der Klis 2000 for a review).  RXTE has
opened up a completely new field of investigation and it is now widely
recognized that fast timing provides one of the most powerful tools
for such objectives ({\em a)} to test General Relativity (GR) in
strong gravity fields, ({\em b)} to constrain the fundamental
properties of collapsed stars (EOS of NS, mass and spin of BH).

The RXTE discoveries were made possible by its large X-ray detector
area ($0.67$ m$^{2}$, sub-$\mu$s time resolution and high data rates.
The RXTE All Sky Monitor (ASM) and the flexibility of operation and
rapid response also played a major role in the success of the mission
(Bradt et al.  1993).

For X-ray timing, the signal to noise ratio at which non-coherent
variability such as a QPO is detected scales as the source count rate
(i.e. the detector area) in the strong source limit (for a photon
counting experiment).  Based on this, here we describe how an
instrument with {\em ten times} the detector area of RXTE could help
to solve some of the outstanding issues in the astrophysics of
collapsed stars.
\section{The science of a future X-ray timing mission}

\begin{figure}[!ht]
\centerline{\includegraphics[width=12pc]{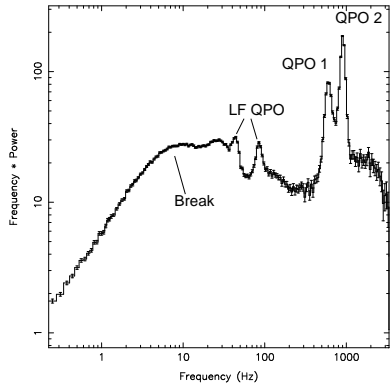}}
     \caption{Power spectrum of Sco\,X-1 obtained with RXTE showing the
     kHz QPOs at $\nu_1$ and $\nu_2$ and the LF QPO (plus its harmonic).
     With EXTRA, about twenty kHz QPO sources will be measured at
     signal-to-noise similar to this (the fainter sources have stronger
     kHz QPOs).  For Sco\,X-1 the signal-to-noise will improve
     tenfold.}
     \label{fig1}
\end{figure}

High-frequency QPOs have been seen in both BHs and NSs.  In BHs, the
QPO models proposed invoke GR effects in the inner accretion disk and
depend strongly on the BH spin, making these QPOs effective probes of
spacetime near the event horizon (see e.g. McClintock 1998).  In NSs,
three types of QPOs are commonly observed (see Fig.  \ref{fig1},
$\nu_{LF}$, 15--60\,Hz, $\nu_1$, 200--800\,Hz and $\nu_2$).  
Evidence is building that there exists a close relationship 
between the QPO properties of NS and BH X-ray binaries. 
In particular a remarkable correlation between the centroid frequency of 
QPOs (or peaked noise components) has been found 
which  extends over nearly 3 decades in frequency and 
encompasses both NS and BH systems 
(Psaltis, Belloni \& van der Klis 1999).
In the
relativistic precession model (Stella \& Vietri 1998), ${\nu_{LF},
\nu_1, \nu_2}$ observed across a wide range of objects are identified
with three fundamental frequencies characterizing the motion of matter
in the strong field of a point mass as predicted by General Relativity
(GR).  The low-frequency QPO at $\nu_{LF}$ is thought to be due to
{\it nodal precession}, dominated by the inertial-frame dragging
predicted by GR in the vicinity of a fast rotating collapsed object.
The lower frequency kHz QPO at $\nu_1$ is identified with relativistic
{\it periastron precession}, an early crucial test of GR. Unlike the
cases of Mercury and the relativistic binary pulsar PSR1913+16, for
which weak field expansions apply, the periastron precession close to
a collapsed star is dictated by strong field effects, requiring a full
general relativistic treatment.  Finally $\nu_2$ is the orbital
(``Keplerian'') frequency; its value alone restricting the allowed
range of mass and radius of the NS (Miller et al.  1998).  Other
models rely on a beat-frequency interpretation for both $\nu_1$ and
$\nu_{LF}$ (Alpar and Shaham, 1985, Miller et al.  1998).

With an order of magnitude better sensitivity, QPOs from a few Hz to
1600 Hz should be detected from many objects with a high enough
significance to use the data for crucial tests.  Regardless of the
physical origin of the QPOs at $\nu_{LF}$ and $\nu_1$, the increased
sensitivity and range will have dramatic benefits.  At higher
frequencies, either strong signatures of the innermost stable circular
orbit (ISCO), which sets an upper bound on $\nu_2$, will be discovered
from several sources (evidence has been found for one source so far:
4U1820-30, Zhang et al.  1998), or the frequencies themselves will
allow the elimination of several candidate equations of state of dense
matter (Miller et al.  1998).  In the {\it relativistic precession}
interpretation, there are several predictions that could be tested.
First, the epicyclic frequency $\Delta\nu=\nu_2-\nu_1$ steeply falls
to zero as $\nu_2$ increases and the orbital radius approaches the
ISCO. The behaviour of the epicyclic frequency in the vicinity of BH
and NS is dominated by strong-field effects and drastically different
from any Newtonian or post-Newtonian approximation.  Hence it provides
a powerful test of the strong field properties of the metric (see
Stella \& Vietri 1999).  According to the model $\Delta\nu$ should
also decrease for low values of $\nu_2$.  Second, $\nu_{LF}$ should
scale as $\nu_2^2$ over a wide range of frequencies (until
``classical'' terms due to stellar oblateness become important).
Observing such scaling would provide an unprecedented test of the
$1/r^3$ radial dependence of $\nu_{LF}$ predicted in the
Lense-Thirring interpretation (Stella \& Vietri 1999).
We note that in the hydrodynamical model of the disk inner boundary
developed by Psaltis \& Norman (2000), the test particle frequencies
(the same as in the relativistic precession model plus a few other
additional frequencies, see also Psaltis 2000) are selected by the
response of the inner disk, when this is subject to a wide-band input
noise.

In the above models, 
the QPO frequencies depend sensitively on the mass
(M) and angular momentum (J) of the compact star, as well as on the
radius at which the QPOs are produced.  M and J could be independently
estimated from waveform measurements at $\nu_{2}$, thus
overdeterminating the problem so that the underlying theories can be
tested in critical ways.  The increased sensitivity will enable QPOs
to be detected within their coherence times.  The cycle waveform,
which it will be possible to reconstruct, depends on the Doppler
shifts associated with the local velocity of the radiating matter in
the emitting blob or spot, as well as on curved-spacetime light
propagation effects.  If the frequency $\nu_2$ of the orbit is known,
QPO waveform fitting yields the mass $M$ (and Kerr spin parameter) of
the compact object.

Nearly coherent oscillations at $\sim$ 300 Hz or $\sim$ 600 Hz have
been observed during type I X-ray bursts from about 10 NS so far (see
Strohmayer et al.  1998 for a review).  These oscillations are
probably caused by rotational modulation of a hot spot on the stellar
surface.  The emission of the hot spot is affected by gravitational
light deflection and Doppler shifts (e.g. Miller \& Lamb 1998).  With
the next generation timing mission, the oscillation will be detected
within one cycle.

\begin{figure}[!t]
\centerline{\includegraphics[width=11pc]{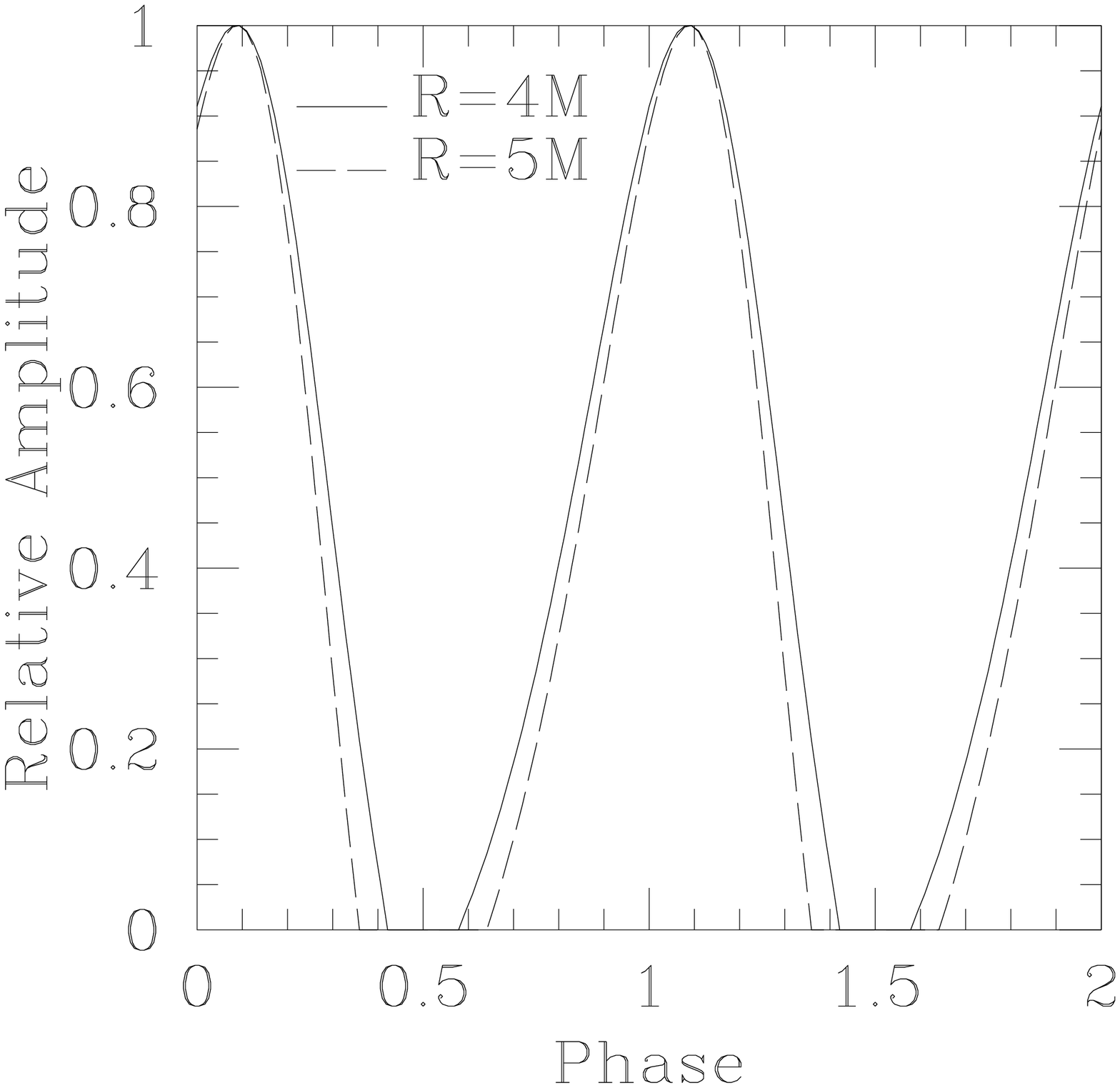}\includegraphics[width=11pc]{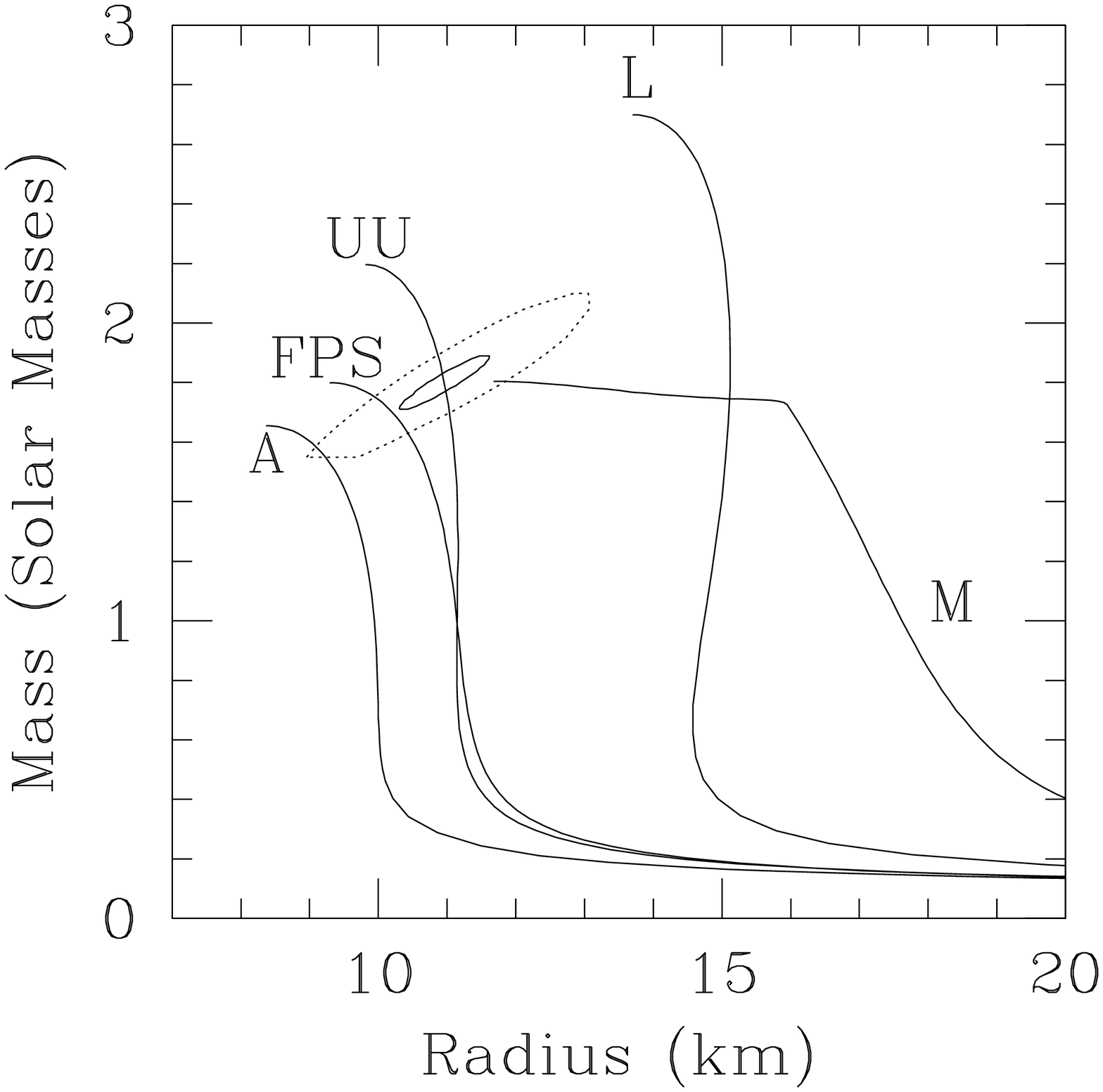}}
   \caption{{\it left)} Simulated cycle waveforms for burst
   oscillations on a $1.8\,M_\odot$ neutron star with a spin frequency
   of 364~Hz of radius $R=4GM/c^2$ (solid line) and radius of
   $R=5GM/c^2$ (dashed line).  The more compact star (with smaller
   $R/M$) has a broader waveform, because it has stronger gravitational
   light deflection.  The less compact star has larger waveform
   asymmetry because its surface rotation velocity is higher.
   Therefore, mass and radius can be determined independently by
   fitting the waveform.  {\it right)} Constraints on mass and radius
   from waveform fitting.  The contours show the 1 $\sigma$ confidence
   regions from RXTE (dotted) and from an instrument 10 times bigger
   than RXTE (drawn).  The mass-radius relations labeled as in Miller
   et al.  (1998) are also shown.  (Courtesy of Cole Miller).}
     \label{fig2}
\end{figure}
The composition and properties of the NS cores have been the subject
of considerable speculation, and remain a major issue in modern
physics : at the highest densities, matter could be composed of pion
or kaon condensates, hyperons, quark matter, or strange matter (see,
e.g., Heiselberg \& Hjorth-Jensen 2000).  By fitting the waveform, it
will be possible to constrain simultaneously the mass and radius of
the star (see fig. \ref{fig2}), and hence determine the equation of 
state of its high
density core.

Beside the few examples described above, with its increased
sensitivity, a future timing mission will also address a broad range
of other astronomical issues (accreting millisecond pulsars,
micro-quasars, X-ray pulsars, CVs, Novae, Soft gamma-ray repeaters,
Anomalous X-ray pulsars, \ldots).  For instance, there is only {\em
one} accreting millisecond pulsar known so far: SAXJ1808-3658
(Wijnands \& Van der Klis 1998, Chakrabarti \& Morgan 1998).  Its
properties suggest that all NS systems should show pulsations at some
level.  In most models, pulse amplitudes cannot be suppressed below
$\sim$0.1\% (rms) without conflicting with spectroscopic or QPO
evidence.  With an instrument 10 times larger than RXTE, the
sensitivity to persistant millisecond pulsations will be well below
this level (pulsations at the level of 0.01\% rms would be detected in
10000 seconds in Sco X-1).  Detection of such pulsations in objects
also showing kHz QPOs and burst oscillations would immediately confirm
or reject several models for these phenomena involving the NS spin
(e.g. Miller et al.  1998).  In addition, it has been suggested that
such objects could be among the brightest gravitational radiation
sources in the sky, emitting a periodic gravitational wave signal at
the star's spin frequency (Bildsten 1998).  Measuring the spin period
very accurately would be therefore of great importance for periodicity
searches with gravitational wave antennas (e.g. Brady et al.  1997).

Similarly for micro-quasars, the link between the very fast disk
transitions observed in X-rays and the acceleration process could be
studied on very short time scales, allowing the non steady state disk
properties and their link to the formation of relativistic jets to be
explored (Belloni et al.  1997, Fender et al.  1999).  This would be
of direct relevance to understanding the properties of AGNs, where
presumably similar jet formation mechanisms operate on a much larger
scale.

\section{Proposed advanced timing missions: EXTRA and RAE}

The ``Experiment for X-ray Timing and Relativistic Astrophysics''
(EXTRA) was proposed to ESA in January 2000 as a F2/F3 flexi-mission.
This mission was not at that stage selected, but studies continue.  In
addition the ``Relativistic Astrophysics Explorer'' (RAE), a closely
related concept, is being discussed in the US (Kaaret et al.  2000).
The objectives and requirements of the two missions are essentially
identical and the two will be discussed together here.

Both EXTRA and RAE have a main instrument consisting of a very large
X-ray detector array, together with an all-sky monitor to provide
source monitoring and alerts to transients.

\begin{figure}
\centerline{\includegraphics[width=22.5pc]{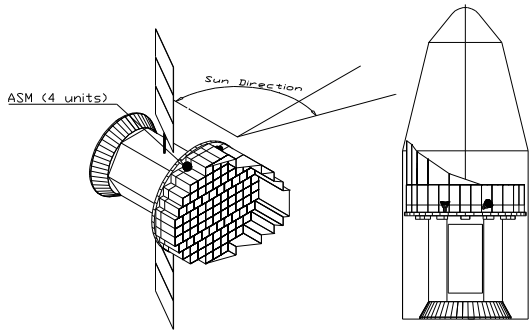}}
    \caption{{\it left)} The EXTRA science payload with the Matra
    Marconi Space LEOSTAR platform.  One of the four ASM modules (WATCH
    design) is shown.  {\it right)} EXTRA in the STARSEM-Soyuz ST
    Ariane 4 type fairing.}
     \label{fig3}
\end{figure}

The main detector array has a geometric area of about 10 m$^2$, giving
about $\sim 6$ m$^2$ effective area at 6--10 keV. An area of this size
can be accommodated within the fairings available with a low-cost
launchers (Starsem Soyuz-ST or Delta II) without the need for
deployable mechanisms (see fig.  \ref{fig3}).  The arrays are highly
modular and in each case Si PIN diodes are being considered for the
main detector.  Si PIN diodes can now be manufactured with high yield
at acceptable cost and provide a superior alternative to the
proportional counters used on XTE. The absence of absorption edges in
the critical 4--6 keV range is important for spectral studies.
Furthermore, they are rugged and require neither high voltages nor
special handling conditions.  Similar Si PIN diodes have been proposed
for the first layer of the ASTRO-E Hard X-ray Detector (Sugizaki et
al.  1997).

By operating the detectors at reduced temperature, needing only
passive cooling through the front face of the detector array, the
leakage current can be reduced to a low level even for an area of
$\sim$ 1 cm$^2$ per element.  The associated capacitance is reduced by
using a thick detector (1--2 mm), also needed for good high energy
response.  In this way good energy resolution can be obtained with an
off-chip pre-amplifier, allowing the silicon and processing to be
optimised for X-ray detection.  Resolution of 600 eV or better can be
expected -- a factor of two better than the RXTE PCA.

For both missions Silicon Drift Detectors (SDDs) are being considered
as a possible alternative to PIN diodes, potentially offering even
better energy resolution and perhaps larger area for a single
detector, so reducing the total number of electronic channels.  SDDs
are, however, at a less advanced stage of development.

The number of electronic channels will be large ($\sim$52416 for
EXTRA) and the resulting electronic complexity requires careful
optimisation and effective use of a modular construction concept.
Similar numbers of channels are, however, being dealt with for the
IBIS instrument on INTEGRAL and on AGILE.

The detector background is reduced because charged particles passing
through the relatively dense detectors will normally deposit an energy
larger than the upper limit of the operating range.  Simulations show
that particle interactions in the collimator and surrounding material
often produce showers of secondaries, so the background can be further
reduced by operating nearby pixels in anti-coincidence with each
other.  Nevertheless, because of the unprecedented area the total
background rate may approach 2500 \ctss\ between 1 and 50 keV.
The missions are intended to study bright sources and the data rates
will often be high, requiring careful optimisation of the on board
electronics and data systems.  With the proposed detector area, the
Crab will produce about 400000 cts/s.  A low, near equatorial, orbit
is to be preferred in order that the background be low and stable and
also because of considerations of data recovery.  The objective is to
telemeter information about the time and energy of every event,
leading to considerable volumes of data.  Nevertheless this turns out
to be feasible with only minimal constraints.  For example, the EXTRA
studies led to the proposal to use the Matra-Marconi Space LEOSTAR
bus, which provides 160 Gbit solid state memory.  By reorienting to
direct a X-band antenna toward the ground station during passes, the
full memory can be dumped during a single pass, using a 240 Mbits
s$^{-1}$ link.  Observations of the brightest sources ($\sim$ 10 Crab)
would thus be restricted to orbits followed by ground station passes.


\section{Conclusions}
The need for a follow-up to RXTE has been clearly identified; fast
timing studies are now widely recognized as a powerful tool to study
the spacetime around collapsed stars, and to derive constraints on the
fundamental properties of neutron stars and black holes.  The studies
carried out for EXTRA have demonstrated that a mission using
solid-state detector technologies providing an order of magnitude
improvement over RXTE can be done within the limits of a modest size
mission.  It has also shown that there are existing platforms with
resources matching the specific needs of such a mission.  More
importantly, the large number of active astronomers that gathered
around the proposals demonstrates that there is a very large community
supportive of such a mission.

It is a real pleasure to thank all the people who have contributed to
the EXTRA proposal (J.L. Atteia, T. Belloni, E. Bravo, A. Brunton, S.
Campana, D. Chakrabarti, F. Cotin, D. Dal Fiume, C. Done, E.C. Ford,
R.P. Fender, C. Hellier, W. Hermsen, G.L. Israel, P. Jean, P. Kaaret,
E. Kuulkers, C. Labanti, S. Larsson, N. Lund, J.E. McClintock, S.
Mereghetti, C.M. Miller, C. Motch, J.F. Olive, M. Orlandini, S.
Paltani, A. Santangelo, R.A. Sunyaev, A. Zdziarski, P. Zycki), and
those of you who have given their scientific or engineering support to
EXTRA (A. Bazzano, L. Burderi, M. B\"oer, H. Bradt, S. Brandt, C.
Castelli, A. Castro-Tirado, M. Cochi, T. Courvoisier, M. Cropper, A.M.
Cruise, D. De Martino, C. Eyles, M. Gilfanov, J.M. Hameury, G. Henry,
E. Kendziorra, W. Kluzniak, L. Kuiper, J.P. Lasota, F.K. Lamb, G. La
Rosa, J. Poutanen, R. Rothschild, R. Svensson, M. Tavani, G. Vedrenne,
N.E. White, R. Wijnands, J. Wilms, W. Zhang).
   
\end{article}

\begin{thebibliography}{}
       \bibitem[\protect\citeauthoryear{}{}]{} Alpar, A. \& Shaham, J.,
       1985, {\em Nature}, 316, 239
       \bibitem[\protect\citeauthoryear{}{}]{} Belloni, T., et al.,
       1997, {\em ApJ}, 479, L145
       \bibitem[\protect\citeauthoryear{}{}]{} Bildsten, L., 1998, {\em
       ApJ}, 501, L89
       \bibitem[\protect\citeauthoryear{}{}]{} Bradt, H., Rothschild, R. \&
Swank, J.H., 1993, {\em A\&AS}, 97, 355
       \bibitem[\protect\citeauthoryear{}{}]{} Brady, P.R., et al.,
       1997, {\em Physics Review}, D57, 2101
\bibitem[\protect\citeauthoryear{}{}]{} Chakrabarti, D. \& Morgan, E.
H., 1998, {\em Nature}, 394, 346
\bibitem[\protect\citeauthoryear{}{}]{} Fender, R. et al., 1999, {\em
MNRAS}, 304, 865
\bibitem[\protect\citeauthoryear{}{}]{} Heiselberg, H. \&
Horth-Jensen, M., 2000, Physics Report, in press
\bibitem[\protect\citeauthoryear{}{}]{} Kaaret, P. et al., 2000, {\em
Astrophysics Letters \& Communications}, in press
\bibitem[\protect\citeauthoryear{}{}]{} Van der Klis, M., 2000,
\newblock {\em ARA\&A}, in press.
\bibitem[\protect\citeauthoryear{}{}]{} McClintock, J.E., 1998, {\em
AIP conference proceeding}, 431, 290
\bibitem[\protect\citeauthoryear{}{}]{} Miller, M.C., Lamb, F.K. \&
Psaltis, D., 1998, {\em ApJ}, 508, 791
\bibitem[\protect\citeauthoryear{}{}]{} Miller, M.C. \& Lamb, F.K. \&,
1998, {\em ApJ}, 499, L39
\bibitem[\protect\citeauthoryear{}{}]{} Psaltis, D., Belloni, T., \&
van der Klis, M. 1999, ApJ, 520, 262
\bibitem[\protect\citeauthoryear{}{}]{} Psaltis, D. \& Norman, C.
2000, ApJ, in press (astro-ph/0001391)
\bibitem[\protect\citeauthoryear{}{}]{} Psaltis, D., 2000, ApJL,
submitted (astro-ph/0010316)
\bibitem[\protect\citeauthoryear{}{}]{} Stella, L. \& Vietri, M.,
1998, \newblock {\em ApJ}, 492, L59

\bibitem[\protect\citeauthoryear{}{}]{} Stella, L. \& Vietri, M.,
1999, \newblock {\em Physical Review
Letters}, 82, 17
\bibitem[\protect\citeauthoryear{}{}]{} Strohmayer,
T., 1998, {\em The Active X-ray Sky}, 69, 129
\bibitem[\protect\citeauthoryear{}{}]{}Sugizaki, M. et al., 1997 ,
{\em Proc.  SPIE} 3115, 244

\bibitem[\protect\citeauthoryear{}{}]{} Wijnands, R. \& Van der Klis,
M., 1998, {\em Nature}, 394, 344
\bibitem[\protect\citeauthoryear{}{}]{} Zhang, W. et al., 1998, {\em
ApJ}, 500, L71
\end{thebibliography}
\end{document}